\documentclass[aps,floats,twocolumn,showpacs,showkeys]{revtex4}
\newcommand{\bastar}{\begin{eqnarray*}}
\newcommand{\eastar}{\end{eqnarray*}}
\newskip\humongous \humongous=0pt plus 1000pt minus 1000pt

\newif\ifdtup

\relax
\newcommand{\be}{\begin{equation}}
\newcommand{\ee}{\end{equation}}
\newcommand{\bea}{\begin{eqnarray}}
\newcommand{\eea}{\end{eqnarray}}

\newcommand{\dfrac}{\displaystyle\frac}
\newcommand{\ba}{\begin{array}}
\newcommand{\ea}{\end{array}}
\newcommand{\nn}{\nonumber}

\begin{document}
\title{On $P_T$-distribution of gluon production rate in constant chromoelectric field}
\author{D. G. Pak}
\email{dmipak@phya.snu.ac.kr}
\affiliation{Center for Theoretical Physics
and School of Physics, \\
College of Natural Sciences,
Seoul  National University, Seoul 151-742, Korea}
\affiliation{Institute of Applied Physics,
Uzbekistan National University, Tashkent, 700-095, Uzbekistan}
\begin{abstract}
A complete expression for the $p_T$-distribution of
the gluon production rate in the homogeneous chromoelectric field
has been obtained. Our result contains a new additional term
proportional to the singular function $\delta(p_T^2)$.
We demonstrate that the presence of this term is consistent with the
dual symmetry of QCD effective action and allows to reproduce the
known result for the total imaginary part of the effective action
after integration over transverse momentum.
\end{abstract}
\pacs{12.38.-t, 12.38.Mh,11.15.-q}
\keywords{quantum chromodynamics, effective action, quark-gluon plasma}
\maketitle

{\bf 1. Introduction}

Recently the calculation of the soft gluon production
rate and its transverse momentum $p_T$-distribution
in a constant chromoelectric field
has been performed \cite{cash,nayak,gavr}.
The motivation for this study is to
describe the particle production in the quark-gluon
plasma produced in hadron collider experiments \cite{exp}.
The obtained result represents a simplest estimate
based on the Schwinger's mechanism of pair creation in the
electric field \cite{schw}
applied to the standard quantum chromodynamics (QCD).

In the present short Letter we give a complete
result for the $p_T$-distribution of the gluon production rate
in homogeneous chromoelectric field which includes a singular term
in addition to the result obtained recently \cite{nayak}.
We demonstrate that our result is consistent with the
analytic structure of the full QCD effective action
(including both real and imaginary parts) and respects the dual symmetry.
Moreover, the additional singular term provides a correct known result
for the total imaginary part of the effective action \cite{ambj1}.
Possible physical implications of the singular
term is discussed.

{\bf 2. Gluon production rate and $p_T$-distribution}

Let us consider the gluon production rate in the constant
color electric field.
For simplicity,
we choose a constant Abelian chromoelectric field
directed along the $z-$axis and defined by the potential
$A_{\mu}^i =-E^i \delta_{\mu 0} z$ (i=3,8).
The integral expression for the one-loop effective Lagrangian
with the transverse momentum $p_T$-dependence
can be written as follows \cite{nayak}
\bea
&\Delta {\cal L} =  \dfrac{g}{16 \pi^3} \lim_{\epsilon\rightarrow0}
\sum_p  \int_{0}^{\infty} d^2 p_T
\dfrac{dt}{t^{1-\epsilon}} \dfrac{E_p}{\sin (gE_p t)} \nn \\
&\Big[\exp(2igE_p t)+\exp(-2igE_p t) \Big]
\exp(-p_T^2 t), \nn \\
& E_p= E^i \vec r_p^i,
\label{eq:nayak}
\eea
where $\vec r_p^i$ (p=1,2,3) are root
vectors of $SU(3)$.
The integration over proper time in (\ref{eq:nayak}) is ill-defined due to
the poles. If we choose the contour above the $t$-axis
starting from $0+\epsilon$ as in quantum electrodynamics (QED)
\cite{yil,ditt} one finds the recent result obtained in \cite{nayak}
\bea
& Im~{\cal L}_{eff} = -\sum\limits_p \sum\limits_{n=1}^\infty
\dfrac{(-1)^n}{n}\dfrac{gE_p}{8 \pi^3}
\exp{(\dfrac{-\pi n p_T^2}{gE_p})}.
\label{eq:nayak2}
\eea
Performing integration over $p_T$ one finds
the total imaginary part
\bea
& Im~{\cal L}_{eff~tot}=\dfrac{}{} \sum_p \dfrac{g^2}{96 \pi} E_p^2, \label{Lsctot}
\eea
which describes the total gluon production
rate. This result had been derived first long time ago \cite{yil}
and it contradicts to results obtained in subsequent papers
\cite{sch,ambj1}. It was shown that the source of this controversy
originated from the unphysical concept of the constant color field in QCD
\cite{ambj2}.

To calculate the $p_T$-distribution of gluon production rate which
is consistent with the total imaginary part of QCD effective action \cite{sch,ambj1}
we will follow the method used in \cite{niel,ambj1}. This
method allows to resolve mainly the ambiguity related
with the pole structure in the integral expression (\ref{eq:nayak}).
Let us start with the functional determinant form
of the one-loop correction $\Delta S$ to the effective action \cite{prd02,jhep}
\bea
\Delta S&=& i\sum_p \Big \{
Tr \ln [Det(-D_p^2 +ig E_p)] \nn \\
&+&Tr \ln [Det(-D_p^2 -ig E_p)] \Big \}.
\eea
\newpage
One can rewrite the last equation in momentum representation
following the method in \cite{ambj1}
\begin{widetext}
\bea
\Delta S= -\dfrac{ig}{8 \pi^3} \sum_p \int d^4x d^2 p_T E_p\Big (
 \ln [p_T^2+igE_p]-\ln [p_T^2-igE_p]
 +2 \sum_{n=1}^\infty \ln[p_T^2+i(2n+1)gE_p] \Big).\label{eqlog}
\eea
\end{widetext}
Using the Schwinger's proper time representation
\bea
& \ln A=-\int_0^\infty \dfrac{dt}{t} \exp[iAt-\epsilon t]
\eea
and performing the series summation one can separate $\Delta S$ into
two parts
\bea
\Delta S&=&\dfrac{ig}{8 \pi^3} \sum_p \int d^4 x d^2 p_T
\dfrac{dt}{t^{1-\epsilon}}  E_p\nn \\
&&(e^{iE_pt}-e^{-iE_pt}) e^{-p_T^2 t} + 2 \Delta S_{sc},  \label{logterm} \\
 \Delta S_{sc}&=& \dfrac{g}{16 \pi^3} \sum_p \int d^4x d^2 p_T
\dfrac{dt }{t^{1-\epsilon}}
      \dfrac{E_p e^{-p_T^2 t}}{\sin (E_p t)}, \label{eqsc}
\eea
where the part $\Delta S_{sc}$ is identical to the expression for the
one-loop effective action of massless scalar QED \cite{schw,ambj1}.
The imaginary part of $\Delta S_{sc}$ is well-known \cite{schw}.
The calculation of the pole
contribution with the Schwinger's
prescription $E_p+i \epsilon$ leads to
the results (\ref{eq:nayak2},\ref{Lsctot}).
However, as we mentioned above, this result is incomplete
because of more complicate analytic structure of the effective action of QCD.
A correct total imaginary part of QCD effective action
with a constant chromoelectric field
had been obtained long time ago in \cite{ambj1}.
One can show that this result can be easily reproduced
if one takes into account the contribution from the pole
at the origin $t=0$. Since the integral in (\ref{logterm})
has no pole at $t=0$ anymore it might seem that the result (\ref{eq:nayak2})
is the only imaginary contribution what we have.
However, the situation is more subtle,
and one should perform a careful analysis of two logarithmic
terms in the last line of the Eqn. (\ref{eqlog})
\bea
\Delta S_1&=&-\sum_p\dfrac{igE_p}{8 \pi^3} \int d^4 x d^2 p_T
\Big ( \ln [p_T^2  \nn \\
&+&ig(E_p - i \sigma) ]
-\ln [p_T^2-ig(E_p - i\sigma)] \Big),
\eea
where, we add an infinitesimal factor $i \sigma$
to define a proper analytic continuation.
To find out a missing part in the $p_T$-distribution
let us consider the derivation of the
total imaginary part in a more detail.
Having integrated over $p_T$ in the last equation
with a cut-off parameter $\Lambda$
one obtains
\begin{widetext}
\bea
\Delta S_1&=&\sum_p\dfrac{igE_p}{8\pi^2}\Big\{
           gE_p\arctan(\dfrac{gE_p}{p_T^2 + \sigma})
+gE_p \Big (\arctan (\dfrac{-gE_p}{p_T^2 - \sigma})
+2 \pi k \Big ) -p_T^2 \ln(p_T^2+igE_p) \nn \\
&&+p_T^2 \ln(p_T^2-igE_p) -igE_p
\ln (p_T^4+g^2E_p^2) \Big\} \Big |_0^{\Lambda^2} . \label{eqamb}
\eea
\end{widetext}
There is still uncertainty related to
the choice of the factors $2 \pi k$ and $\sigma$
which defines a possible analytic continuation of the
logarithmic function and
chromoelectric field.
Let us consider two choices: \\
(A) the choice $\sigma>0$, $k=0$ leads to the imaginary part
\bea
&&Im~{\cal L}_{1}=-\sum_p \dfrac{g^2 E_p^2}{8 \pi} .\label{first}
\eea
Notice, the prescription $E_p-i\sigma$ is the same as that used
in studies of the QCD vacuum stability
based on causality requirement
\cite{prd02,cho99};\\
(B) the adopting $\sigma=0$ , $k=1$ at the
lower integration limit and neglecting
the contribution at the upper limit, i.e., $k=0$ at $p_T=\Lambda^2$,
leads to
\bea
&&Im~{\cal L}_{2}=-\sum_p \dfrac{g^2E_p^2}{4 \pi}.\label{imL1}
\eea
The corresponding expressions for the
total imaginary parts for these two cases
(including the contribution
(\ref{Lsctot}) from $\Delta S_{sc}$) are the following
\bea
&& Im~{\cal L}_{1tot}=-\sum_p \dfrac{11 g^2E_p^2}{96 \pi}, \nn \\
&& Im~{\cal L}_{2tot}=-\sum_p \dfrac{23 g^2E_p^2}{96 \pi}. \label{im12}
\eea
For the case of $SU(2)$ QCD
the first result, ${\rm Im}~{\cal L}_{1tot}$  (with no summation
over $p$ in the above equations), corresponds to a perturbative contribution
which has been obtained by perturbative methods \cite{sch,jhep},
while the second result,  ${\rm Im}~{\cal L}_{2tot}$, represents a total contribution
derived first by Ambjorn and Hughes \cite{ambj1}.
In both results the sign is negative, that means
one has gluon pair annihilation instead of gluon
creation \cite{sch}.
Notice that the imaginary part ${\rm Im}~{\cal L}_{1tot}$ is dual
symmetric to the vanishing imaginary part for a possible
stable chromomagnetic vacuum \cite{prd02,jhep},
whereas the Ambjorn-Hughes result is
dual symmetric to the magnetic imaginary part of Nielsen-Olesen
for a constant external chromomagnetic field, i.e., for a different
physical problem. From the
dual symmetry consideration
we should assign the Ambjorn-Hughes type result ${\rm Im}~{\cal L}_{2tot}$
as a correct one for the case
of external constant chromoelectric field.

The key point in our derivation of (\ref{im12}) is that
the non-vanishing values of
the imaginary  parts (\ref{first},\ref{imL1}) originate from the
behavior near vicinity of the point $p_T^2=0$.
In the first case (A)
the imaginary part (\ref{first}) is not zero because
the sum of angle functions on the right hand side of Eqn. (\ref{eqamb})
does not vanish in the infinitesimal region $p_T^2<\sigma$.
In the second case (B)
the angle argument $2 \pi k$ changes its
value from $0$ at the upper integration limit to
$2 \pi$ at the lower limit $p_T^2=0$
and must have behavior like a step function since
$k$ is integer.
This implies that in both cases
the corresponding $p_T$-distribution is represented by
a singular $\delta$-function.
We choose a numeric factor in front of the
$\delta$-function which reproduces
the result (\ref{imL1}) in agreement with
Ambjorn-Hughes result \cite{ambj1}
\bea
&& Im~{\cal L}_{2}(p_T)= - \delta(p_T^2)
\sum_p \dfrac{g^2E_p^2}{4 \pi}. \label{eq:delta}
\eea
The appearance of the singular function
$\delta(p_T^2)$
reflects the low-momentum phenomenon of
strong interaction.
A simple heuristic argument
is that any color field in the confinement phase should
be confined, so that the constant field
configuration must shrink and be localized
in a small space region.
One should stress again that the constant color
field is not a well-defined physical concept
and the imaginary part of the effective action
depends on a concrete physical problem
\cite{ambj2}.  Remind, that the negative imaginary part
implies the gluon pair annihilation
which assumes the presence of a source supporting the existence of
the constant field itself \cite{sch}.

{\bf 3. Consistence with dual symmetry}

An important feature of the effective action
is its invariance under the dual transformation
$H_p \rightarrow -i E_p$, $E_p \rightarrow i H_p$.
For a general constant background the dual symmetry of $SU(2)$ QCD
had been established in \cite{cho99}.
The duality provides a powerful tool to check the consistency of
obtained results. In particular, since
the duality is intrinsically based on the analytic structure of full
effective action, the real and imaginary parts must be interelated
in a non-trivial manner.
From this point of view, the consistence
with duality should imply the existence of a singular term in the
chromomagnetic counterpart of the $p_T$-distribution (\ref{eq:nayak}).
Besides, such a term should reproduce the well-known Nielsen-Olesen imaginary part
of the effective action after integration over transverse momentum.

Let us consider a formal expression for a dual symmetric magnetic counter-part to Eqn. (\ref{eq:nayak})
\bea
\Delta {\cal L} &=&  \dfrac{g}{16 \pi^3}
\sum_p  \int_{0}^{\infty} d^2 p_T
\dfrac{dt}{t^{1-\epsilon}} \dfrac{H_p}{\sinh (gH_p t)}\nn \\
&& \times \Big(e^{2gH_p t}+e^{-2gH_p t} \Big)
e^{-p_T^2 t}. \label{dualmag}
\eea
Notice that for the case of magnetic background
the parameter $p_T$  should be treated
just as a formal variable as it follows from the
derivation method of the Eqn. (\ref{eq:nayak})
\cite{nayak,itz}.
The integral in (\ref{dualmag}) has a strong infra-red divergency
when $p_T^2<gH_p$ which causes an imaginary part.
Using the series expansion
\bea
\dfrac{x}{\sinh x}=1-\dfrac{x^2}{6}-\dfrac{2x^4}{\pi^2}\sum_{n=1}^\infty
\dfrac{(-1)^n}{n^2} \dfrac{1}{x^2+n^2 \pi^2}
\eea
one can perform the integration over proper time
with the $\zeta-$function regularization. Finally,
using analytical properties of the logarithmic
and cosine integral functions, (notice that $H_p$ has an infinitesimal
factor $+i \epsilon$ due to the dual symmetry \cite{cho99})
one can find that the imaginary part is non-vanishing
only when $p_T^2 < gH_p$
\bea
Im&\Delta&\!\!\!{\cal L}=\dfrac{1}{16\pi^2}\sum \limits_p (2gH_p-p_T^2)  \nn \\
   && -\dfrac{g}{8\pi^3}
\sum\limits_{p,n=1}^\infty \dfrac{(-1)^{n}}{n} H_p
 \sin  \big(\dfrac{\pi n p_T^2}{gH_p}\big).
\eea
Surprisingly,
using the identity
\bea
&&\sum\limits_{n=1}^\infty \dfrac{(-1)^n}{n} \sin (n \pi x)\equiv\nn \\
 &&~~ \dfrac{i}{2}(\ln[1+e^{i \pi x}]-\ln[1+e^{-i \pi x}]),
\eea
one can check
that inside the interval $0<p_T^2<g H_p$
all $p_T^2$ dependent terms
are mutually cancelled,
so that, the imaginary
part has a very simple $p_T$ dependence like a step function
\bea
Im\,\Delta {\cal L}=
\left\{ {\sum\limits_p \dfrac{g H_p}{8 \pi^2}~,~~~~0 \leq p_T^2 \leq gH_p ,
\atop  0~,~~~~~~~~~~~~p_T^2 \geq gH_p .}\right .
\eea
This result is a dual counter-part to the singular term
(\ref{eq:delta}). Performing integration over $p_T$ we can easily
reproduce the total imaginary part of Nielsen-Olesen \cite{niel}.

{\bf 4. Conclusion}

There has been a lot of controversies on
the imaginary part of the effective
action with a constant field background
\cite{ambj1,yil,ditt,sch,ambj2,niel,prd02,jhep}.
The fact that the concept of constant color field
itself is not well-defined and requires a specification
of the source \cite{sch} and a concrete physical problem \cite{ambj2}
leads to the uncertainty in calculating the imaginary part
of the effective action.
Additional controversies originate from the several sources.
The first derivation of the positive imaginary part of the effective action
by Yildiz and Cox \cite{yil}
includes only the pole contributions obtained by using the Schwinger's prescription
for passing poles. The negative value of the imaginary part
obtained first by Schanbacher \cite{sch} was provided by perturbative calculation.
This result has been confirmed later by applying a slightly
different perturbative calculation scheme \cite{jhep}. The physical meaning
of the negative imaginary part of the effective action is closely related with
the asymptotic freedom and corresponds to gluon pair annihilation \cite{sch}.
A complete total contribution to the imaginary part of the
effective action including all non-perturbative contributions
had been calculated by Ambjorn and Hughes \cite{ambj1}.

In conclusion, we have obtained a complete strict expression for the $p_T$-distribution
of gluon production rate in a constant chromoelectric field. We have shown that an additional
singular term appears in $p_T$-distribution. This result might be of
a pure academic interest because it is hardly possible to create a constant chromoelectric field
in experiments. Nevertheless, it would be very interesting to find possible physical
implications of such singular term, for instance, in
quark-gluon plasma, where the concept of constant color field
can be adopted in mean field approximation.

{\bf Acknowledgements}

Author acknowledges Profs. Y.M. Cho, N.I. Kochelev and S.P. Gavrilov
for useful discussions. The work is supported in part by the ABRL Program of
Korea Science and Engineering Foundation (R14-2003-012-01002-0).


\begin{thebibliography}{99}
\bibitem{cash} A. Casher, H. Neuberger, and S. Nussinov, Phys. Rev.
{\bf D20}, 179 (1979).
\bibitem{nayak} G. C. Nayak and  P. van Nieuwenhuizen,
Phys.Rev. {\bf D71} 125001 (2005); G. C. Nayak, Phys.Rev. {\bf D72}, 125010 (2005).
\bibitem{gavr} S.P. Gavrilov, D.M. Gitman and J.L. Tomazelli, hep-th/0612064.
\bibitem{exp} Y. Schutz, J. Phys. {\bf G30}, S903 (2004);
L. Mclerran and M. Gyulassy, Nucl. Phys. {\bf A750}, 30 (2005);
B. Muller, nucl-th/0508062.
\bibitem{schw} J. Schwinger, Phys. Rev. {\bf 82}, 664 (1951).
\bibitem{ambj1} J. Ambjorn and R. J. Hughes, Phys. Lett. {\bf B113}, 305 (1982).
\bibitem{yil} A. Yildiz and P.H. Cox, Phys. Rev. {\bf D21}, 1095 (1980);
M. Claudson, A. Yildiz, and P.H. Cox, Phys. Rev. {\bf D22}, 2022 (1980).
\bibitem{ditt} W. Dittrich and M. Reuter, Phys. Lett. {\bf B128}, 321, (1983);
S.L. Adler, Phys. Rev. {\bf D23}, 2905 (1981).
\bibitem{sch} V. Schanbacher, Phys. Rev. {\bf D26}, 489 (1982).
\bibitem{ambj2} J. Ambjorn and R. J. Hughes, Nucl. Phys. {\bf B197}, 113 (1982).
\bibitem{niel} N. K. Nielsen and P. Olesen,
Nucl. Phys. {\bf B144}, 376 (1978); H. B. Nielsen and M. Ninomiya,
Nucl. Phys. {\bf B156}, 1 (1979).
\bibitem{prd02} Y.M. Cho, H.W. Lee, and D.G. Pak,
Phys. Lett. {\bf B 525}, 347 (2002); Y. M. Cho and D. G. Pak,
Phys. Rev. {\bf D65}, 074027 (2002).
\bibitem{jhep} Y.M. Cho, D. G. Pak, and M. Walker,
JHEP {\bf 05}, 073 (2004); Y. M. Cho and M. L. Walker,
Mod. Phys. Lett. {\bf A19}, 2707 (2004).
\bibitem{cho99} Y.M. Cho and D. G. Pak,  Procs. of TMU-YALE
Symp. on "Dynamics of Gauge Fields"
(Univ. Acad. Press, Tokyo, 1999); hep-th/0006051.
\bibitem{itz} C. Itzikson and J-B. Zuber, {\it Quantum Field Theory} (McGraw-Hill) 1985, p.193.
\end{thebibliography}
\end{document}